\documentclass{iau} 
\usepackage{graphicx}

\title[HELP project - a dreamed-of multiwavelength
dataset for SED fitting] 
{HELP project - a dreamed-of multiwavelength dataset for SED fitting: the influence of used models for the main physical properties of galaxies.}

\author[Katarzyna Ma\l{}ek \& the HELP team]   
{Katarzyna Ma\l{}ek$^{1,2}$, 
  Veronique Buat$^2$, 
  Denis Burgarella$^2$, 
  Yannick  Roehlly$^{2,3}$, 
  Raphael Shirley$^4$  
 \and the HELP team}

\affiliation{$^1$National Centre for Nuclear Research, {ul.Pasteura~7, 02-093} Warszawa, Poland, \\ email: {\tt katarzyna.malek@ncbj.gov.pl} \\[\affilskip]
$^2$Aix Marseille Univ. CNRS, CNES, LAM Marseille, France,\\
$^3$Univ Lyon, Univ Lyon1, ENS de Lyon, CNRS, Centre de Recherche Astrophysique de Lyon UMR5574, F-69230, Saint-Genis-Laval, France\\
$^4$Astronomy Centre, Department of Physics and Astronomy, University of Sussex, Falmer, Brighton BN1 9QH, UK}

\pubyear{2018}
\volume{341}  
\jname{Title of your IAU Symposium}
\editors{A.C. Editor, B.D. Editor \& C.E. Editor, eds.}
\begin{document}

\maketitle

\begin{abstract}
The Herschel Extragalactic Legacy Project (HELP) focuses to  publish an astronomical multiwavelength catalogue of millions of objects over 1300~deg$^2$ of the Herschel Space Observatory survey fields.  Millions of galaxies with ultraviolet--far infrared photometry {make} HELP a perfect sample for testing spectral energy distribution fitting models, and to prepare tools for next-generation data.  
In the frame of HELP collaboration we estimated the main physical properties of all galaxies from the HELP database and we checked a new procedure to select peculiar galaxies from large galaxy sample and we investigated the influence of used modules for stellar mass estimation. 

\end{abstract}

\firstsection 
\section{Introduction}

The primary objective of the Herschel Extragalactic Legacy Project (HELP project, Oliver et al., in preparation, \cite[Vaccari 2018]{Vacdustatt_modified_CF00cari:2016}) founded by FP7 European Union  is to provide homogeneously calibrated multiwavelength catalogues covering roughly 1300~deg$^2$ of the extragalactic Herschel Space Observatory surveys  (HSO, \cite[{Pilbratt} \etal\ 2010]{Pilbratt2010})  at wide redshift range. 
Millions of galaxies with good coverage of ultraviolet--far infrared spectral range {make} HELP a perfect sample to prepare tools for next-generation data.  
The detailed description of a final master list creation of 170 million objects, selected at 0.36---4.5 $\mu$m from HSO, depth maps etc.  can be found in Shirley et al., 2019 MNRAS (under review). 
The catalogues supported by spectroscopic (if possible) or photometric redshift (\cite[{Duncan} \etal\ 2018]{Duncan2017}) will allow for colour-colour/colour-flux analysis, multi-wavelength spectral energy distribution (SED) fitting and many more statistical studies  of the low-to-intermediate redshift galaxy population formation and evolution over cosmic time. 

Tab.~\ref{tab1} shows the list of the HSO fields used for HELP project. 
It demonstrates that HELP not only created a huge multiwavelength, homogenized  database, but also focuses both on wide and deep fields, with different area on the sky. 
This careful selection and the final data product can remove the barriers to multiwavelength data studies on the statistical level.

\begin{table}
  \begin{center}
  \caption{Overview of 23 fields  used for HELP project.}
  \label{tab1}
 {\scriptsize
  \begin{tabular}{|l|r|r|}\hline 
{\bf HELP field name} & {\bf number of objects} & {\bf area [deg$^2$]} \\ \hline
AKARI-NEP          & 531 746    & 9.2   \\
AKARI-SEP          & 844 172    & 8.7   \\
Bootes             & 3 367 490  & 11    \\
CDFS-SWIRE         & 2 171 051  & 13    \\
COSMOS             & 2 599 374  & 5.1   \\
EGS                & 1 412 613  & 3.6   \\
ELAIS-N1           & 4 026 292  & 14    \\
ELAIS-N2           & 1 783 240  & 9.2   \\
ELAIS-S1           & 1 655 564  & 9.0   \\
GAMA-09            & 12 937 982 & 62    \\
GAMA-12            & 12 369 415 & 63    \\
GAMA-15            & 14 232 880 & 62    \\
HDF-N              & 130 679    & 0.67  \\
Herschel-Stripe-82 & 50 196 455 & 363   \\
Lockman-SWIRE      & 4 366 298  & 22    \\
HATLAS-NGP         & 6 759 591  & 178   \\
SA13               & 9 799      & 0.27  \\
HATLAS-SGP         & 29 790 690 & 295   \\
SPIRE-NEP          & 2 674      & 0.13  \\
SSDF               & 12 661 903 & 111   \\
xFLS               & 977 148    & 7.4   \\
XMM-13hr           & 38 629     & 0.76  \\
XMM-LSS            & 8 704 751  & 22    \\
\hline
Total:             & 171 570 436 & 1270 \\ \hline
  \end{tabular}
  }
 \end{center}
\end{table}

\section{Data and short overview of the method}

The  European Large Area ISO Survey North~1 (ELAIS~N1, 13.51~deg$^2$ area centred at 16$^{h}$10$^{m}$01$^{s}$ +54$^{o}$30$^{'}$36$^{''}$, \cite[{Oliver} {et~al.} 2000]{Oliver2000}) was a pilot field for HELP. 
The HELP homogenized catalogue of ELAIS~N1 includes 50~135 galaxies with good ultraviolet (UV)--far infrared (IR) measurements (quality criterion requires at least two optical -- near IR measurements and at least two  two of five Herschel measurements with signal to noise ratio $\geqq$2). 
We used the sample of 50~135 galaxies and we  estimated the key physical parameters (i.e.~stellar mass, star formation rate, dust luminosity) by fitting SED to all of them using  Code Investigating GALaxy Emission (CIGALE, \cite[Burgarella et~al. 2005]{Burgarella2005}, \cite[Noll et~al. 2009]{Noll2009}, and \cite[Boquien et~al. 2018]{Boquien2018}).

CIGALE is designed to estimate the physical parameters  by comparing modelled galaxy SEDs to observed ones. 
CIGALE conserves the energy balance  between the dust-absorbed stellar emission and its re-emission in the IR. 
A more detailed description of the code can be found in \cite{Boquien2018}.

\begin{figure}[t]
 \vspace*{-0.2 cm}
\begin{center}
 \includegraphics[width=3.0in]{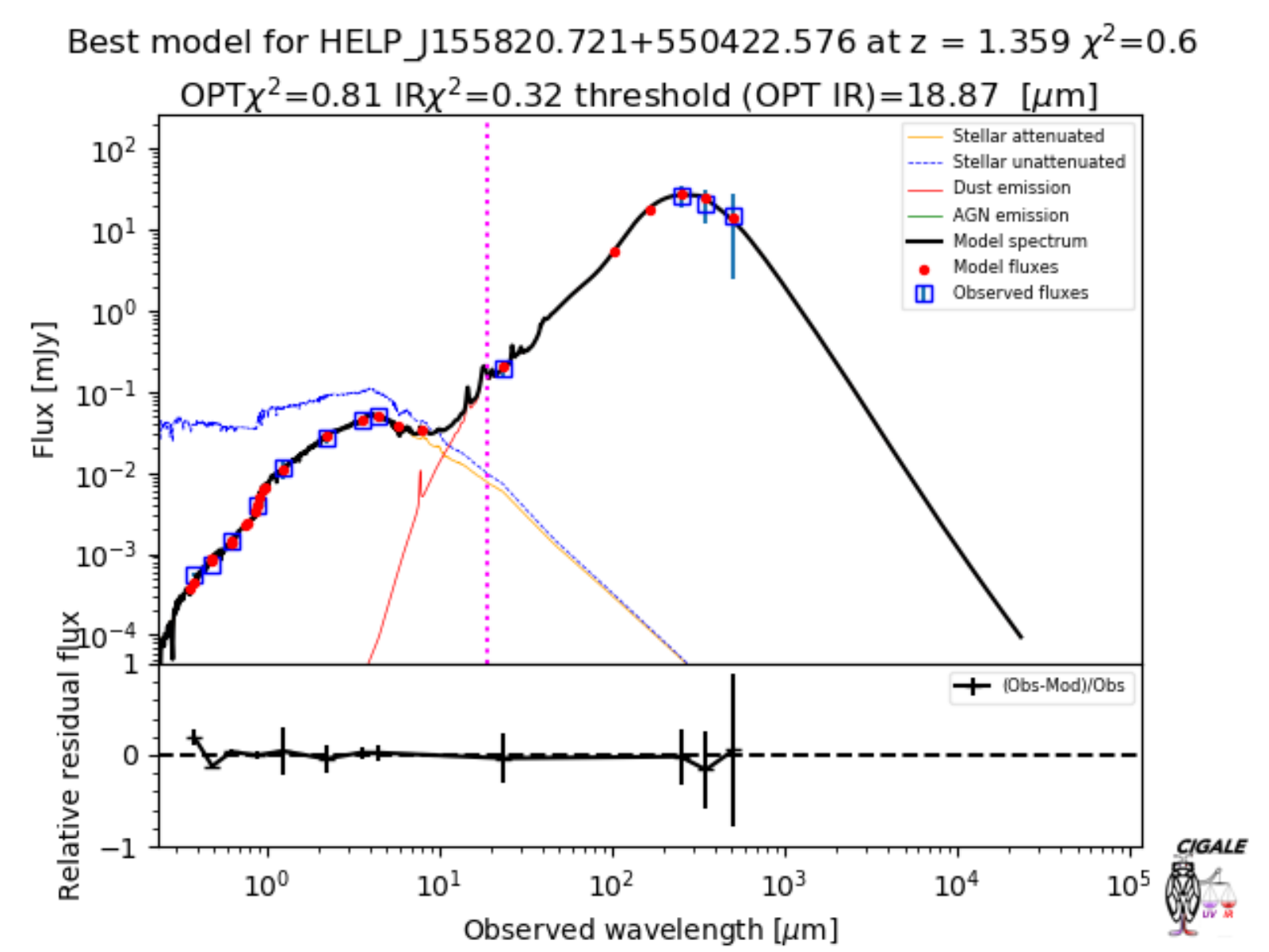} 
 \caption{An example of SED fitting result. Open squares represent observed fluxes, while filled circles correspond to the model fluxes. The final model is plotted as a solid black line.  The relative residual fluxes are plotted at the bottom of the spectra.}
   \label{fig1}
\end{center}
\end{figure}

All adopted parameters used for modules are presented in Table.~\ref{tab2}. 
More detailed discussion of used parameters and description of addition quality tests for SED fitting procedure for ELAIS~N1 field can be found in \cite{Malek18}. 
An exemplary fit of SED, showing typical photometric coverage of the spectra is shown in Fig.~\ref{fig1}. 

\begin{table}
  \begin{center}
  \caption{Main modules and input parameters used in CIGALE for the analysis of the high-z sample. The first column lists the CIGALE model, the second provides a brief description of the main parameters, and the third one shows the range of the selected values. }
  \label{tab2}
 {\scriptsize
  \begin{tabular}{|l|r|r|}\hline 
{\bf CIGALE module} & {\bf main parameter} & {\bf description} \\ \hline 
SFH delayed + additional  burst      & $\tau$ of the main stellar population model [Myr]    & 3~000   \\
 &  $\tau$ of the late starburst population model [Myr]    & 10~000   \\
 &  mass fraction of the late burst population    & 0.001--0.300   \\
SSP:  \cite{bruzual03}  &  initial mass function & \cite{Chabrier2003IMF}   \\ 
dust attenuation: \cite{CF00} & $\rm A_V$ in the BCs & 0.3--3.8\\
 &  power law slopes (BC and ISM) & -0.7\\
dust emission \cite{draine07} & minimum radiation field ($\rm U_{min}$) & 5.0, 10.0, 25.0\\
& mass fraction of PAH &  1.12, 2.5, 3.19\\
& power law slope dU/dM  ($U^{\alpha}$) &  2.0, 2.8 \\
AGN emission: \cite{fritz06} & fractional contribution of AGN & 0.0, 0.15, 0.25, 0.8\\
\hline  
  \end{tabular}
  }
   \end{center}
\end{table}

\section{Impact of the dust attenuation law on the stellar mass}

Based on the statistically significant sample of $\sim$50~000 galaxies we check the influence of different dust attenuation recipes on the main physical parameters calculated for all HELP galaxies; stellar mass, star formation rate and dust luminosity. 
We perform the SED fitting of ELAIS~N1 galaxies by assuming three different dust attenuation laws separately: \cite{CF00}, widely used in the literature \cite{calzetti00}, and \cite{LoFaro2017} -- dust attenuation recipe created in the framework of the HELP project for Ultra Luminous Infrared Galaxies at redshift $>$ 2. 
This test allows us to analyze the impact  of the assumed law on estimated physical parameters.
We find that the attenuation law has an important impact on the stellar mass estimation (on average leading to disparities of a factor of 2), and we derived the relation between stellar mass estimates obtained by those three different attenuation laws. 
Found recipes (published in \cite[Ma{\l}el \etal\ 2018]{Malek18}) can  help to homogenize estimated stellar masses from different attenuation laws, and allow to make more precise comparisons, sample selection or study of so called main sequence (stellar mass versus star formation rate relation) between different SED fitting procedures.

\begin{figure}[t]
\begin{center}
 \includegraphics[width=3.4in]{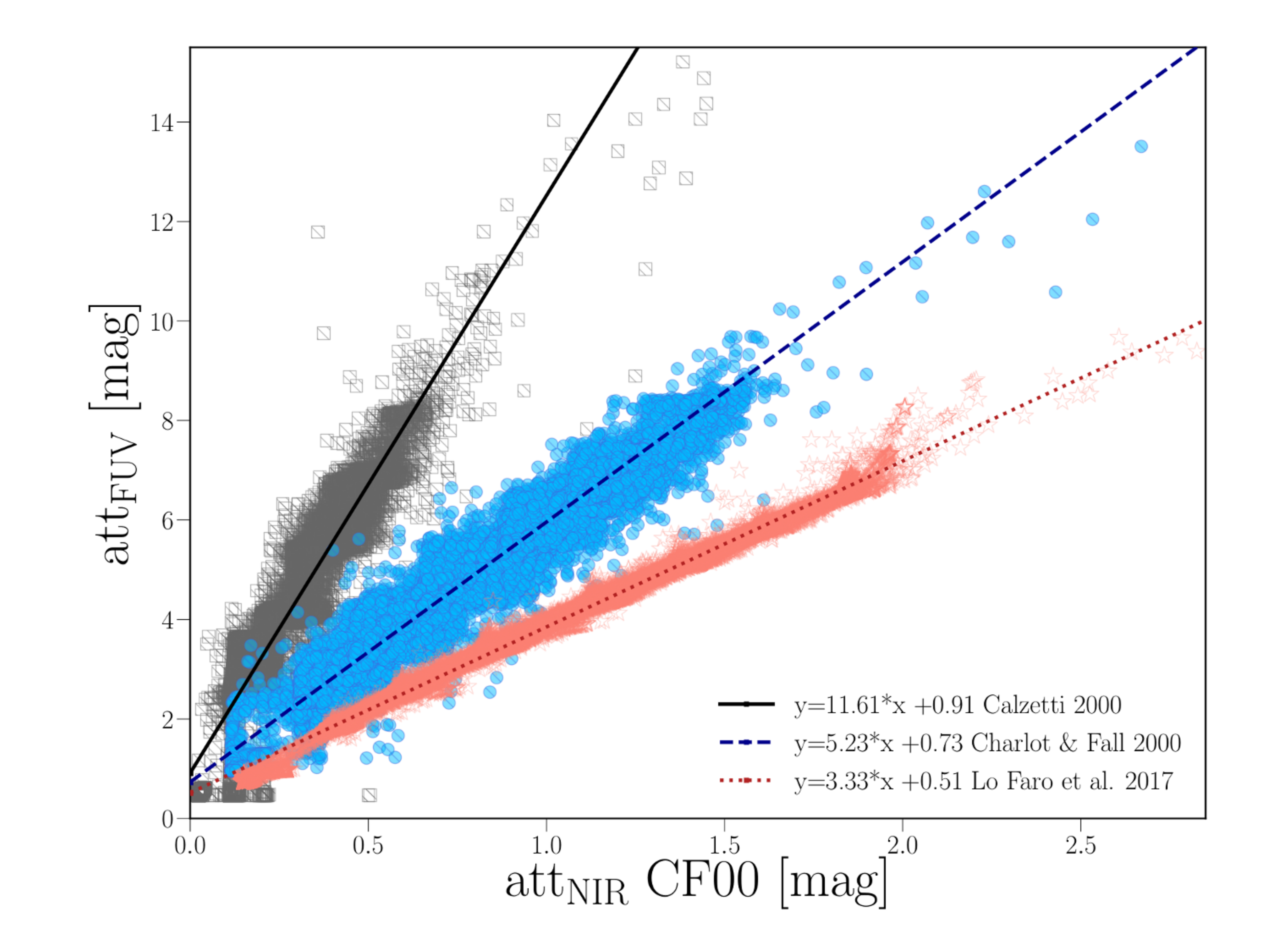} 
 \caption{Relation between attenuation in near infrared band and attenuation in ultaviolet band for all three laws used in the analysis. Open black squares represent \cite{calzetti00} recipe, blue dots -- \cite{CF00} law, and orange stars correspond to the \cite{LoFaro2017} law.}
   \label{fig2}
\end{center}
\end{figure}

We check that the differences in obtained stellar masses are closely related to the shape of each attenuation law at near IR wavelengths. 
Fig.~\ref{fig2} shows relation between attenuation in near IR band and far UV band for all three attenuation laws used in our analysis. 
This figure presents that the range and distribution of attenuation in ultraviolet band is similar for \cite{CF00}, \cite{calzetti00}, and \cite{LoFaro2017}, however the attenuation obtained in near infrared band is meaningly different. 
Similar result, showing that Calzetti recipe leads to steeper slopes, not consistent with radiation transfer models results, was found by \cite{Buat2018} based on the infrared complete sample  of galaxies in the COSMOS 3D-HST CANDELS field at  0.6$<$z$<$1.6.  
Similar impact of the attenuation law on the stellar mass was found by \cite{Mitchell2013} based on the semi-analytic galaxy formation model
GALFORM (\cite[Cole \etal\ 2000]{Cole2000}).

\end{document}